\documentclass[prb,aps,twocolumn,showpacs,nobibnotes,epsf]{revtex4}

\usepackage{graphicx}
\usepackage{dcolumn}
\usepackage{bm}
\usepackage{SIunits}

\begin{document}
\title{Physical properties of $A_x$Fe$_{2-y}$S$_2$ ($A$=K, Rb and Cs) single crystals}
\author{J. J. Ying, Z. J. Xiang, Z. Y. Li, Y. J. Yan, M. Zhang, A. F. Wang, X. G. Luo and X. H. Chen}
\altaffiliation{Corresponding author} \email{chenxh@ustc.edu.cn}
\affiliation{Hefei National Laboratory for Physical Science at
Microscale and Department of Physics, University of Science and
Technology of China, Hefei, Anhui 230026, People's Republic of
China\\}

\begin{abstract}
We successfully synthesized two new compounds Rb$_x$Fe$_{2-y}$S$_2$
and Cs$_x$Fe$_{2-y}$S$_2$ which were isostructural with
K$_x$Fe$_{2-y}$Se$_2$ superconductor. We systematically investigated
the resistivity, magnetism and thermoelectric power of
$A_x$Fe$_{2-y}$S$_2$ ($A$=K, Rb and Cs) single crystals. High
temperature resistivity and magnetic measurements show anomalies
above 500 K depending on $A$ which are similar to
$A_x$Fe$_{2-y}$Se$_2$. Discrepancy between ZFC and FC curves was
observed in K$_x$Fe$_{2-y}$S$_2$ and Rb$_x$Fe$_{2-y}$S$_2$, while it
disappears in Cs$_x$Fe$_{2-y}$S$_2$. Our results indicate the
similar magnetism between $A_x$Fe$_{2-y}$S$_2$ and
$A_x$Fe$_{2-y}$Se$_2$ at high temperature.
\end{abstract}


\vskip 300 pt

\maketitle

The discovery of iron-based high temperature superconductors
attracted much attention in these years and it provided a new family
of materials to explore the mechanism of high-$T_c$
superconductivity besides cuprate
superconductors\cite{Kamihara,chenxh, ren, rotter}. Among all the
iron-based superconductors, the anti-PbO type FeSe$_x$ owning the
simplest structure with the edge-sharing FeSe$_4$ tetrahedra formed
FeSe layers stacking along the {\sl c}-axis. FeSe$_x$ displays a
lower $T_{\rm c}$ of 8 K at ambient pressure \cite{MKWu} and $T_{c}$
can reach 37 K (onset) under 4.5 GPa.\cite{cava} The corresponding
pressure dependent ratio of $T_{\rm c}$ can reach as large as
d$T_{\rm c}$/d$P$ of $\sim$ 9.1 K/GPa, which is the highest among
all the Fe-base superconductors.\cite{cava, Mizuguchi2, Margadonna,
Garbarino}. Recently, by intercalating K, Rb, Cs and Tl between the
FeSe layers, superconductivity has been enhanced to about 30 K
without any external pressure in Fe-Se
system\cite{xlchen,Mizuguchi,Wang, Ying, Krzton, Fang}. The
$A_x$Fe$_{2-y}$Se$_2$ ($A$=K, Rb, Cs and Tl) superconductors are
isostructural to 122 iron-pnictide superconductors. Unlike the
iron-pnictide superconductors, large amount of $A$ and Fe vacancies
exist in the samples and it is amazing that superconductivity can
still survive in the Fe layers. $A_x$Fe$_{2-y}$Se$_2$ undergoes
structural transition and antiferromagnetic transition around 500
K\cite{Baowei,Liu} which is much higher than that in iron-pnictide
superconductors. The mechanism of superconductivity and its
relationship with structural and antiferromagnetic transition in
this system are still unknown. Superconductivity can be gradually
suppressed through S doping\cite{Guo, Lei1}. For
K$_x$Fe$_{2-y}$S$_2$, it is a small gap semiconductor and shows
spin-glass behavior below 32 K\cite{Lei2}. The fundamental
differences between the superconducting samples
K$_x$Fe$_{2-y}$Se$_2$ and non-superconducting samples
K$_x$Fe$_{2-y}$S$_2$ is not clear and detailed physical properties
of intercalated FeS samples need further investigation.

In this paper, we successfully synthesized two new compounds
Rb$_x$Fe$_{2-y}$S$_2$ and Cs$_x$Fe$_{2-y}$S$_2$ which were
isostructural to K$_x$Fe$_{2-y}$S$_2$. We systematically
investigated the physical properties of $A_x$Fe$_{2-y}$S$_2$ ($A$=K,
Rb and Cs) single crystals. Resistivity and magnetic measurements
show anomalies above 500 K depending on $A$. We attribute these
anomalies to the structural and antiferromagnetic transitions
compared to the isostructural superconducting $A_x$Fe$_{2-y}$Se$_2$
($A$=K, Rb and Cs) compounds. K$_x$Fe$_{2-y}$S$_2$ shows
semiconductor behavior at low temperature the same to the previous
report. The resistivity of Rb$_x$Fe$_{2-y}$S$_2$ and
Cs$_x$Fe$_{2-y}$S$_2$ shows broad humps which is similar to
superconducting $A_x$Fe$_{2-y}$Se$_2$ samples, but no
superconducting transition was observed in these samples. Single
crystals of $A_x$Fe$_2$S$_2$ were characterized by powder X-ray
diffraction (XRD), X-ray single crystal diffraction,
Energy-dispersive X-ray spectroscopy (EDX), direct current (dc)
magnetic susceptibility, electrical transport and thermoelectric
power(TEP) measurements. Resistivity below 400 K was measured using
the Quantum Design PPMS-9. The resistivity measurement above 400 K
was carried out with an alternative current resistance bridge
(LR700P) by using the a Type-K Chromel-Alumel thermocouples as
thermometer in a home-built vacuum resistance oven. Magnetic
susceptibility was measured using the Quantum Design SQUID-MPMS. A
high-temperature oven was used in the SQUID-MPMS for magnetic
susceptibility measurement above 400 K.

Single crystals $A_x$Fe$_{2-y}$S$_2$ were grown by Bridgeman method
with the nominal composition $A$:Fe:S = 0.8:2:2. Starting material
FeS was obtained by reacting Fe powder with S powder with Fe: S = 1:
1 at 700$\celsius$ for 4 hours. Alkali metals and FeS powder were
sealed into two wall quartz tubes. The mixture was heated to 1050
$\celsius$ in 4 hours and then kept at this temperature for 2 hours,
and later slowly cooled down to 750 $\celsius$ at a rate of 6
$\celsius$/ hour. After that, the temperature was cooled down to
room temperature by shutting down the furnace. Plate-like single
crystals can be cleaved from the final products.
\begin{figure}[t]
\centering
\includegraphics[width=0.5\textwidth]{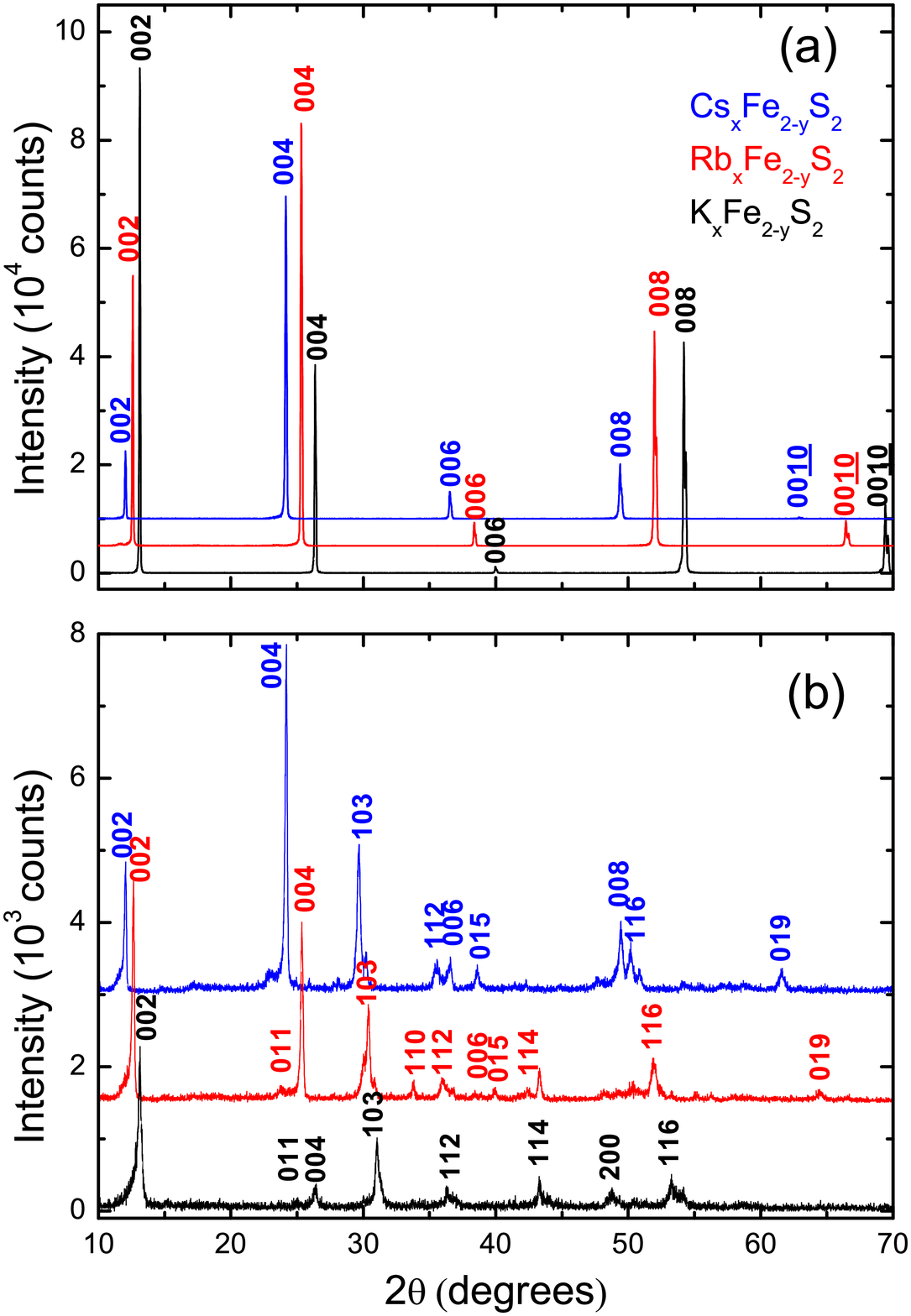}
\caption{(color online). (a) The single crystal x-ray diffraction
pattern of $A_x$Fe$_{2-y}$S$_2$, Only (00$l$) diffraction peaks show
up which indicating that the c axis is perpendicular to the plane of
the plate. (b) X-ray diffraction pattern of the powdered
$A_x$Fe$_{2-y}$S$_2$.} \label{fig1}
\end{figure}

Figure 1 shows the X-ray single crystal diffraction and powder XRD
after crushing the single crystals to powder for
$A_x$Fe$_{2-y}$S$_2$. Only (00$l$) diffraction peaks were observed,
suggesting that the crystallographic c axis is perpendicular to the
plane of the single crystal. The lattice constants of c-axis for
K$_x$Fe$_{2-y}$S$_2$, Rb$_x$Fe$_{2-y}$S$_2$ and
Cs$_x$Fe$_{2-y}$S$_2$ were determined to be 13.546 ${\rm \AA}$,
14.070 ${\rm \AA}$ and 14.801 ${\rm \AA}$ respectively, it is
consistent with the increasing of alkali ion radius from K to Cs.
The lattice constants of a-axis for K$_x$Fe$_{2-y}$S$_2$,
Rb$_x$Fe$_{2-y}$S$_2$ and Cs$_x$Fe$_{2-y}$S$_2$ were determined to
be 3.772 ${\rm \AA}$, 3.789 ${\rm \AA}$ and 3.824 ${\rm \AA}$
respectively. The actual compositions of K$_x$Fe$_{2-y}$S$_2$,
Rb$_x$Fe$_{2-y}$S$_2$, Cs$_x$Fe$_{2-y}$S$_2$ were determined by EDX
to be K$_{0.68}$Fe$_{1.70}$S$_2$, Rb$_{0.74}$Fe$_{1.67}$S$_2$ and
Cs$_{0.72}$Fe$_{1.71}$S$_2$ respectively.
\begin{figure}[t]
\centering
\includegraphics[width=0.5\textwidth]{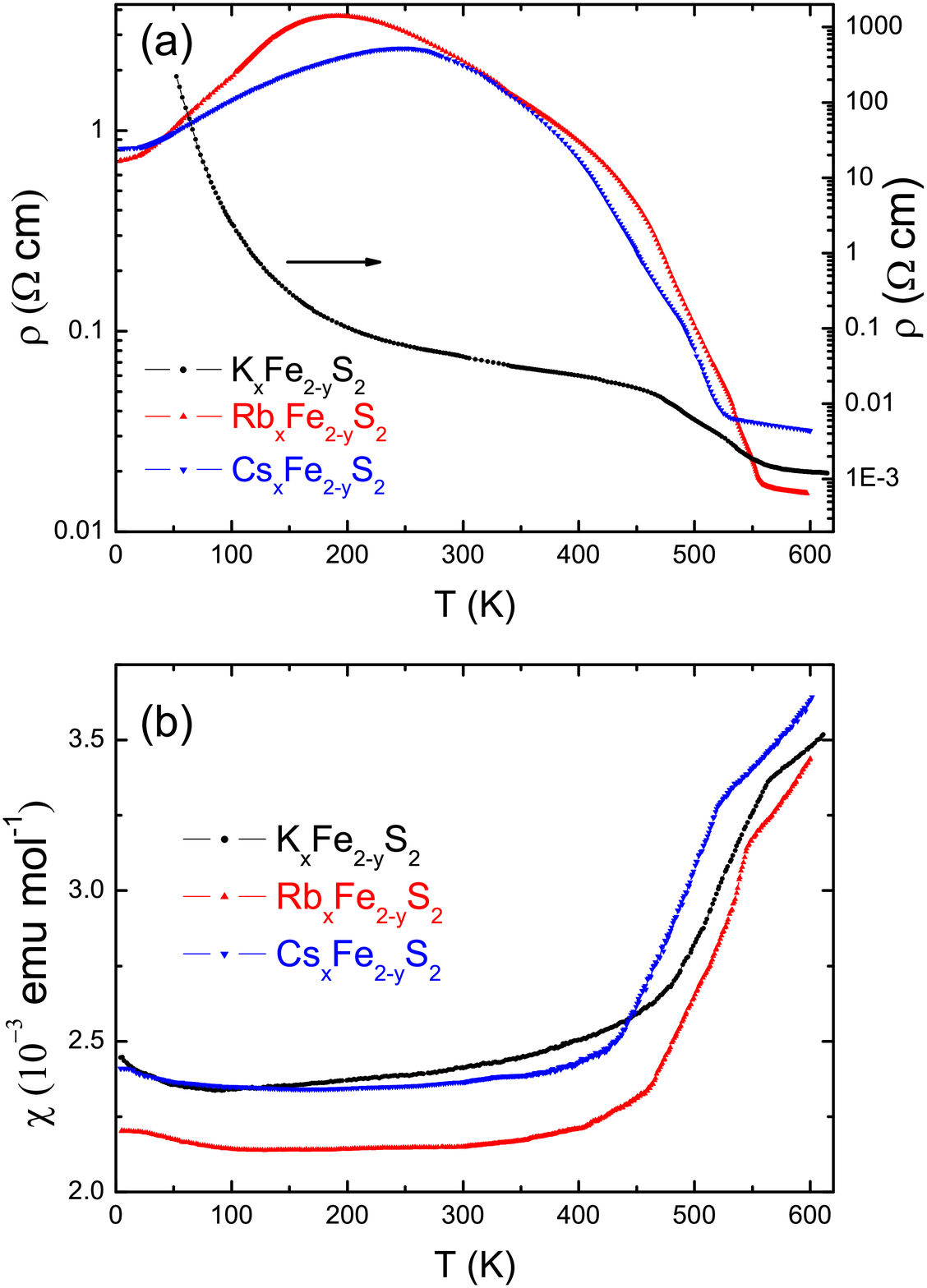}
\caption{(color online). (a) The temperature dependence of the
resistivity for $A_x$Fe$_{2-y}$S$_2$. (b) The temperature dependence
of the magnetic susceptibility with field of 5 T applied along the
ab plane for $A_x$Fe$_{2-y}$S$_2$.} \label{fig2}
\end{figure}

Figure 2(a) shows the temperature dependence of the resistivity in
the temperature ranges from 4 K to 620 K for $A_x$Fe$_{2-y}$S$_2$.
K$_x$Fe$_{2-y}$S$_2$ shows semiconductor behavior at low
temperature, being similar to the previous reports\cite{Guo, Lei1}.
The resistivity of Rb$_x$Fe$_{2-y}$S$_2$ and Cs$_x$Fe$_{2-y}$S$_2$
shows broad humps which are similar to superconducting samples
$A_x$Fe$_{2-y}$Se$_2$\cite{Luo, Ying2}, however no superconducting
transition was observed at low temperature. The hump temperatures
$T_{hump}$ for Rb$_x$Fe$_{2-y}$S$_2$ and Cs$_x$Fe$_{2-y}$S$_2$ are
180 K and 250 K respectively. Sharp decrease in resistivity can be
observed above 500 K for all the samples with increasing the
temperature. The anomalies of resistivity for all the samples
clearly indicate phase transition in this kind of materials above
500 K. In order to deep investigate this phase transition, we take
magnetic measurement with the temperature up to 610 K. Fig. 2(b)
shows the temperature dependence of magnetic susceptibility with the
field magnitude of 5 T applied within ab plane for
$A_x$Fe$_{2-y}$S$_2$. The magnetic susceptibilities show sharp drop
above 500 K with decreasing the temperature for all the samples
which indicates antiferromagnetic transition at high temperature in
this kind of materials. At the temperature below 400 K, the
susceptibilities nearly show no temperature dependence.
\begin{figure}[t]
\centering
\includegraphics[width=0.5\textwidth]{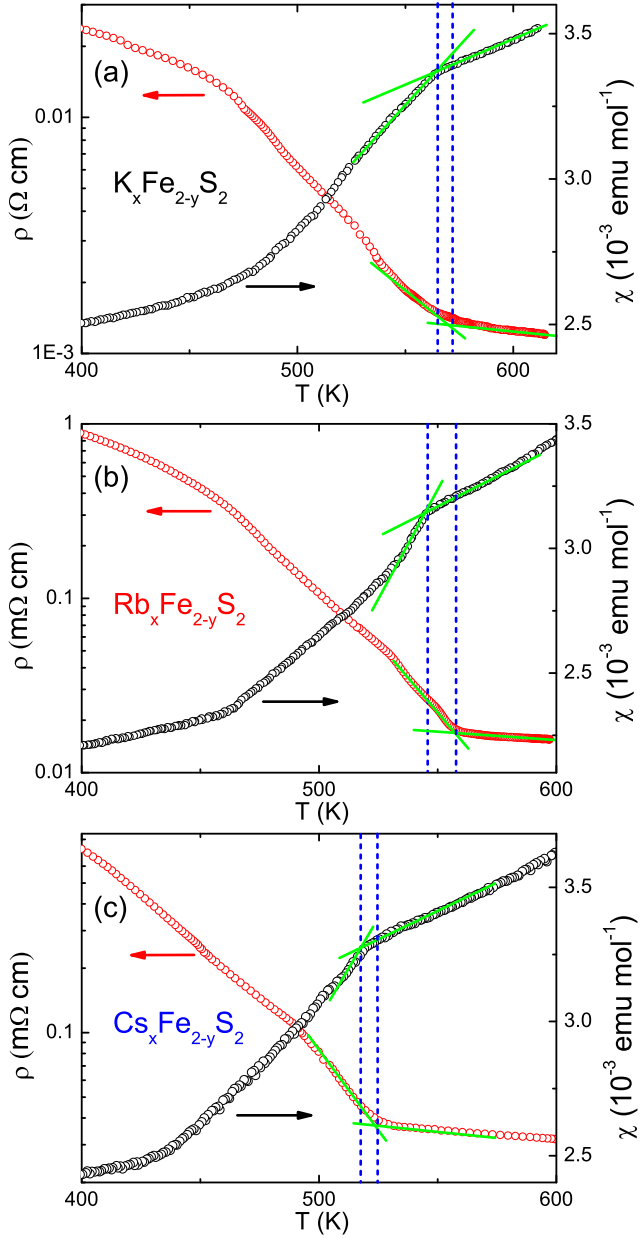}
\caption{(color online). The temperature dependence of resistivity
and susceptibility around the temperature of $T_N$ for
K$_x$Fe$_{2-y}$Se$_2$ (a), Rb$_x$Fe$_{2-y}$Se$_2$ (b) and
Cs$_x$Fe$_{2-y}$Se$_2$ (c).} \label{fig3}
\end{figure}

We compared the high temperature resistivity with the susceptibility
above 400 K as shown in Fig. 3. We found the anomaly temperature
determined by the resistivity ($T_r$) is slightly higher than the
antiferromagnetic transition temperature ($T_N$). The detailed
temperatures of $T_r$ and $T_N$ for $A_x$Fe$_{2-y}$S$_2$ are listed
in Table I. $T_r$ and $T_N$ increase with decreasing the lattice
constant of c-axis. The temperature dependence of resistivity and
susceptibility of $A_x$Fe$_{2-y}$S$_2$ is very similar to its
isostructural $A_x$Fe$_{2-y}$Se$_2$ superconducting samples. $T_r$
and $T_N$ in $A_x$Fe$_{2-y}$S$_2$ are about 20 K higher than those
in $A_x$Fe$_{2-y}$Se$_2$\cite{Liu}. The structures between
$A_x$Fe$_{2-y}$S$_2$ and $A_x$Fe$_{2-y}$Se$_2$ are the same and
similar Fe vacancy ordering was observed in
K$_x$Fe$_{2-y}$S$_2$\cite{Lei1}. Considering the similarities of the
structure and physical properties between $A_x$Fe$_{2-y}$S$_2$ and
$A_x$Fe$_{2-y}$Se$_2$, we ascribe the sudden increase of resistivity
to the structural transition arising from the Fe vacancy ordering.
We can conclude that all the samples of $A_x$Fe$_{2-y}$$Ch_2$($A$=K,
Rb and Cs, $Ch$=S,Se) exhibit the common features of
antiferromagnetic and structural transition at high temperature.
Although $A_x$Fe$_{2-y}$S$_2$ and $A_x$Fe$_{2-y}$Se$_2$ show the
same physical properties in normal state, no superconductivity was
found in $A_x$Fe$_{2-y}$S$_2$.
\begin{figure}[t]
\centering
\includegraphics[width=0.5\textwidth]{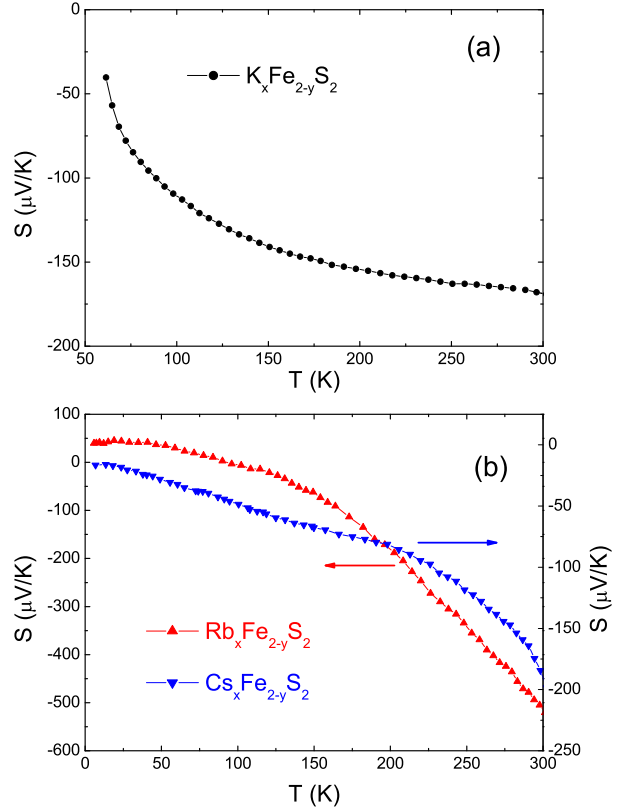}
\caption{(color online). The temperature dependence of
thermoelectric power of K$_x$Fe$_{2-y}$S$_2$ (a),
Rb$_x$Fe$_{2-y}$S$_2$ and Cs$_x$Fe$_{2-y}$S$_2$ (b).} \label{fig4}
\end{figure}

\begin{table*}[t]
\tabcolsep 0pt \caption{The summary of the actual compositions
determined by EDX analysis with errors within 5$\%$, the a-axis
lattice parameters and the c-axis lattice parameters, the abnormal
temperature determined by the susceptibility (T$_N$) and resistivity
(T$_r$) for all the crystals $A_{x}$Fe$_{2-y}$S$_2$ ($A$ = K, Rb,
Cs).} \vspace*{-12pt}
\begin{center}
\def\temptablewidth{1\textwidth}
{\rule{\temptablewidth}{1pt}}
\begin{tabular*}{\temptablewidth}{@{\extracolsep{\fill}}cccccccc}
 sample name & $A$: Fe: Se & a (${\rm \AA}$) & c (${\rm \AA}$)&$T_N$ (K) &$T_r$ (K) \\   \hline
      K$_x$Fe$_{2-y}$S$_2$ & 0.68: 1.70: 2 &3.772  & 13.546 & 565 & 571 \\
      Rb$_x$Fe$_{2-y}$S$_2$ & 0.74: 1.67: 2 &3.789 & 14.070 & 546 & 558 \\
      Cs$_x$Fe$_{2-y}$S$_2$ & 0.72: 1.71: 2 &3.824 & 14.801 & 518 & 524 \\
       \end{tabular*}
       {\rule{\temptablewidth}{1pt}}
\end{center}
\end{table*}
Figure 4(a) shows the temperature dependence of TEP for
K$_x$Fe$_{2-y}$S$_2$. The TEP for K$_x$Fe$_{2-y}$S$_2$ exhibits
large negative value at room temperature. With decreasing the
temperature, the absolute value of TEP gradually decreases and
became small negative value at low temperature which is similar to
K$_x$Fe$_{2-y}$Se$_{2-y}$S$_y$\cite{Kefeng Wang}. Fig. 4(b) shows
the temperature dependence of TEP for Rb$_x$Fe$_{2-y}$S$_2$ and
Cs$_x$Fe$_{2-y}$S$_2$. The TEP for both the two samples shows
negative value at room temperature. With decreasing the temperature,
the absolute value of TEP for Rb$_x$Fe$_{2-y}$S$_2$ gradually
decreases and TEP shows small positive value at low temperature
which indicates the multiple-band electronic structure in
Rb$_x$Fe$_{2-y}$S$_2$. For Cs$_x$Fe$_{2-y}$S$_2$, the absolute value
of TEP decreases with decreasing the temperature and still exhibit
negative at low temperature. The different temperature dependence of
TEP for these samples might be due to the different doping contents
which was caused by the different $A$ and Fe content in the samples.
\begin{figure}[t]
\centering
\includegraphics[width=0.5\textwidth]{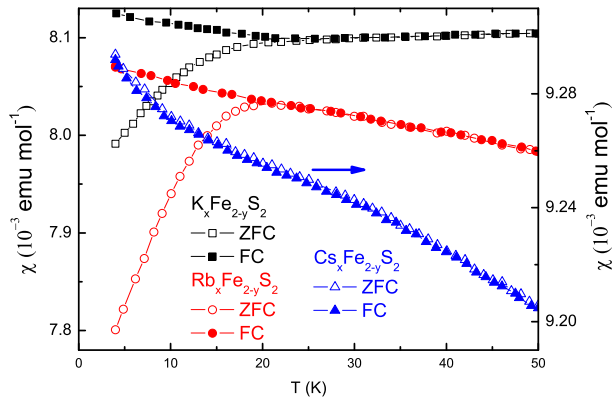}
\caption{(color online). ZFC and FC dc magnetic susceptibility with
the field 1000 Oe applied within the ab plane for
$A_x$Fe$_{2-y}$S$_2$ at low temperature.} \label{fig5}
\end{figure}

Fig.5 shows ZFC and FC dc magnetic susceptibility with the field
1000 Oe applied within the ab plane for $A_x$Fe$_{2-y}$S$_2$ at low
temperature. Obvious discrepancy between ZFC and FC curves was found
for K$_x$Fe$_{2-y}$S$_2$ and Rb$_x$Fe$_{2-y}$S$_2$ below 20 K. Such
behavior is not yet understood. However, discrepancy between ZFC and
FC curves was not observed in Cs$_x$Fe$_{2-y}$S$_2$.

The physical properties of $A_x$Fe$_{2-y}$S$_2$ are very similar to
its isostructural compounds $A_x$Fe$_{2-y}$Se$_2$ except for the
absence of superconductivity. The absence of superconductivity in
$A_x$Fe$_{2-y}$S$_2$ might be closely related to the structure and
the content of $A$ and Fe\cite{Lei1}, however, it is still an open
question and other experiment need to elucidate this problem. The
superconductivity can be gradually suppressed through S substitution
for Se in K$_x$Fe$_{2-y}$Se$_2$\cite{Lei1}. However, the
antiferromagnetic transition at high temperature stays the same
which indicates the local magnetic exchange interactions between
$A_x$Fe$_{2-y}$S$_2$ and $A_x$Fe$_{2-y}$Se$_2$ are almost the same.
The transition temperatures for $A_x$Fe$_{2-y}$$Ch_2$ slightly
increase with the c-axis lattice parameters decreasing. The
antiferromagnetic transition seems not directly connected to the
superconductivity in this system. Almost the same antiferromagnetic
transition temperatures between superconducting samples and
insulting samples of K$_x$Fe$_{2-y}$Se$_2$ samples also prove
this\cite{Yan}. The different resistivity behaviors at low
temperature for these samples might be due to the phase separation
in this system which is similar to K$_x$Fe$_{2-y}$Se$_2$
samples\cite{Fei}. The metallic phase without superconductivity may
coexists with the insulating phase in this system, so with the
different composition of metallic phase and insulating phase may
results in the different hump temperatures. The maximum resistivity
of Rb$_x$Fe$_{2-y}$S$_2$ and Cs$_x$Fe$_{2-y}$S$_2$ is about one
order larger than the normal state resistivity of superconducting
$A_x$Fe$_{2-y}$Se$_2$\cite{Ying}, even in K$_x$Fe$_{2-y}$S$_2$ no
metallic resistivity behavior was observed. It indicates that the
proportion of metallic phase in $A_x$Fe$_{2-y}$S$_2$ is smaller than
that in $A_x$Fe$_{2-y}$Se$_2$. TEP clearly indicate the multi-band
electronic structure in this system and the electronic structure is
strongly dependent on $A$ and Fe content in the samples. Discrepancy
between ZFC and FC curves at low temperature might be related to the
$A$ and Fe content, or to the different Fe arrangement. The origin
for the appearance of such behavior in this system still needs to be
investigated.

In conclusion, we had successfully synthesized two new compounds
Rb$_x$Fe$_{2-y}$S$_2$ and Cs$_x$Fe$_{2-y}$S$_2$ which were
isostructural with K$_x$Fe$_{2-y}$S$_2$. Resistivity and
susceptibility anomalies were observed above 500 K for
$A_x$Fe$_{2-y}$S$_2$ single crystals which was similar to
superconducting $A_x$Fe$_{2-y}$Se$_2$ samples. TEP indicated the
multi-band electronic structure which was common in iron-based
superconductors. Discrepancy between ZFC and FC curves at low
temperature was possibly related to the $A$ and Fe content, or to
the local environment of Fe ions\cite{Lei1}.

{\bf ACKNOWLEDGEMENT} This work is supported by the National Basic
Research Program of China (973 Program, Grant No. 2012CB922002 and
No. 2011CB00101), National Natural Science Foundation of China
(Grant No. 11190020 and No. 51021091), the Ministry of Science and
Technology of China, and Chinese Academy of Sciences.
\\

\end{document}